\documentclass[preprint,12pt,authoryear]{elsarticle}
\usepackage{amssymb}
\usepackage{gensymb}
\usepackage{lineno}

\usepackage{xcolor}
\usepackage{soul} 
\usepackage{ulem}
\usepackage{enumitem}

\begin{document}

\begin{frontmatter}

\title{How hidden 3D structure within crack fronts reveals energy balance}

\author[label1]{Meng Wang}

\author[label2]{Mokhtar Adda-Bedia}

\author[label3]{John M. Kolinski}

\author[label1]{Jay Fineberg\corref{cor1}}

\affiliation[label1]{organization={The Racah Institute of Physics},
            addressline={The Hebrew University of Jerusalem}, 
            city={Jerusalem},
            postcode={91904}, 
            country={Israel}}

\affiliation[label2]{organization={Université de Lyon, Ecole Normale Supérieure de Lyon},
            addressline={CNRS, Laboratoire de Physique}, 
            city={Lyon},
            postcode={F-69342}, 
            country={France}}
            
\affiliation[label3]{organization={Engineering Mechanics of Soft Interfaces},
            addressline={School of Engineering, Ecole Polytechnique F\'ed\'erale de Lausanne}, 
            city={Lausanne},
            postcode={1015}, 
            country={Switzerland}}

\cortext[cor1]{\ead{jay@mail.huji.ac.il}}

\begin{abstract}

Griffith's energetic criterion, or `energy balance', has for a century formed the basis for fracture mechanics; the energy flowing into a crack front is precisely balanced by the dissipation (fracture energy) at the front. If the crack front structure is {\it not} properly accounted for, energy balance will either appear to fail or lead to unrealistic results. Here, we study the influence of the secondary structure of low-speed crack propagation in hydrogels under tensile loading conditions. We first show that these cracks are bistable; either simple (cracks having no secondary structure) or faceted crack states (formed by steps propagating along crack fronts) can be generated under identical loading conditions. The selection of either crack state is determined by the form of the initial `seed' crack; perfect seed cracks generate simple cracks while a small local mode~III component generates crack fronts having multiple steps. Step coarsening eventually leads to single steps that propagate along crack fronts. As they evolve, steps locally change the instantaneous structure and motion of the crack front, breaking transverse translational invariance. In contrast to simple cracks, faceted cracks can, therefore, no longer be considered as existing in a quasi-2D system. For both simple and faceted cracks we simultaneously measure the energy flux and local dissipation along these crack fronts over velocities, $v$, spanning $0<v<0.2c_R$ ($c_R$ is the Rayleigh wave speed). We find that, in the presence of secondary structure within the crack front, the implementation of energy balance must be generalized for 3D systems; faceted cracks reveal energy balance, only when we account for the local dynamic dissipation at each point along the crack front. 

\end{abstract}

\begin{keyword}
Griffith criterion \sep Energy balance \sep Bistability \sep Fracture toughness \sep Faceted crack
\end{keyword}

\end{frontmatter}

\section{Introduction}

The existence of cracks causes a significant decrease in the practical strengths of materials compared to theoretical values. Once propagating, cracks are the vehicle that drives material failure. Crack initiation and propagation are of crucial importance in questions ranging from the stability of materials \citep{freund1998dynamic, sun2012highly, bouchbinder2014dynamics, ducrot2014toughening, yang2019polyacrylamide} to earthquake nucleation and dynamics \citep{rosakis2002intersonic,svetlizky2014classical,gvirtzman2021nucleation}. It is therefore surprising that most of our detailed theoretical knowledge of fracture is generally limited to ideal systems. In this paper, we take a closer look at a central tenet of fracture: energy balance of non-ideal systems. In particular, we will examine the validity of energy balance in the presence of cracks having non-trivial internal structure.

Important progress on the dynamics of `simple' cracks has been made in two-dimensional and quasi-two-dimensional systems \citep{freund1998dynamic, bouchbinder2014dynamics, long2021fracture}. We refer to `simple cracks' as those with no secondary structure. Simple cracks are, conceptually, simple branch cuts having a $r^{-1/2}$ singularity, where $r$ is the distance from the crack tip. Upon propagation, they form a clean `mirror' surface in their wake. Linear elastic fracture mechanics (LEFM) provides the basis of our understanding of simple cracks \citep{freund1998dynamic}. LEFM assumes linear elastic material response, except in the process zone, a small region surrounding a crack's tip where all dissipative and nonlinear processes take place. Outside the process zone, LEFM predicts a singular stress field which is characterized by a $K/\sqrt{r}$ singularity, where $K$ is the stress intensity factor that quantifies the amplitude of the stress field. $K$, depending on the applied loading and geometrical configuration of the crack system, determines how the crack tip will behave in the given system.

\subsection{Simple and not so simple cracks}

Simple cracks, however, do not necessarily possess a `simple' structure. Over the past decade or so, studies have found that the classic square-root singularity at the tip of a crack may break down as the large strains near a crack's tip force the surrounding material to become nonlinearly elastic. We will still refer to cracks, however, as simple cracks so long as no instabilities develop and the crack front forms, in its wake, a trivial mirror-like surface.

At sufficiently high speeds, simple cracks do become unstable. The nonlinear elastic region will drive oscillatory cracks that generate wavy crack paths \citep{bouchbinder2009dynamic, chen2017instability, vasudevan2021oscillatory}. In brittle materials, rapid cracks may also lose stability in other ways. Beyond a critical velocity of $\sim$0.3-0.4$c_{R}$ (where $c_{R}$ is the material's Rayleigh wave speed), mode~I cracks can lose stability to micro-branches \citep{ravi1984experimentalII,ravi1984experimentalIII,sharon1996microbranching,sharon1999confirming, katzav2007theory}, where the simple main crack spontaneously sprouts daughter cracks; microscopic cracks that extend away from the main crack until arresting. 

Recent work has shown that even very slow, nearly quasistatic simple cracks may also become unstable. A small mode~III component is sufficient to cause simple cracks to break up into discrete segments separated by sharp propagating steps. As these cracks propagate across a crack front, they leave in their wake segmented, faceted fracture surfaces \citep{Tanaka.98,lazarus2008comparison, baumberger2008magic, pham2014further}. Phase-field modeling has shown that a planar crack can indeed become faceted \citep{pons2010helical}, when $K_{III} / K_{I}$ crosses a material-dependent threshold \citep{leblond2011theoretical}. Initially planar (simple) cracks then evolve into segmented arrays that evolve from a nonlinear helical instability. Once crack segmenting takes place through this mechanism, experiments have shown that steps will merge and the segmented fracture surfaces will coarsen in a self-similar way \citep{ronsin2014crack, chen2015crack}. Recent experiments in polyacrylamide hydrogels \citep{kolvin2018topological} both revealed how step topology leads to their stability and that local symmetry breaking causes the steps to propagate along the crack front. These studies also revealed that steps have a complex local 3D structure. Obviously, when a simple crack develops such secondary structures, it can no longer be considered as a 1D object having a point-like singularity at its tip, but a 3D object bounded by 1D crack fronts. This internal 3D structure \citep{kolvin2017nonlinear} significantly alters the local in-plane dynamics of the crack front.

In many materials, {\it simple} crack propagation at very low speeds seems to be unreachable. In crystalline materials \citep{thomson1971lattice, marder1993instability} the lattice trapping effect prevents a simple crack from propagating at very low speeds and jumps to cracks propagating at finite speeds are expected to result. Velocity jumps that preclude slow crack speeds have been observed in experiments in {\it amorphous} materials, where lattice trapping should not play a role. Examples include rubber-like materials \citep{morishita2016velocity, kubo2021dynamic} (possibly due to a dynamic rubbery-glassy transition at slow speeds), PMMA \citep{fineberg_instability_1992}, and soft brittle hydrogels \citep{livne2005universality}. In hydrogels such as  polyacrylamide elastomers, stable simple crack propagation has never been observed at low crack speeds \citep{kolvin2018topological, cao2018porous} and simple cracks are generally observed to jump to $v\sim 0.2c_R$. In polyacrylamide gels, steps may form when crack fronts are locally perturbed, but it has never been clear if a fundamental reason exists for why slow simple cracks have never been observed at low crack velocities.   

\subsection{Energy Balance in 2D and 3D systems}

A central tenet of fracture mechanics is that the motion of a crack is governed by energy balance. \citet{griffith1921vi} suggested energy balance as a criterion for a crack's extension, where the energy flux into the crack tip, $G$ is balanced by the fracture energy $\Gamma$, the energy dissipated per unit crack extension. $G$, the energy release rate, is a quadratic function of $K$ \citep{irwin1957analysis,freund1998dynamic}, and $\Gamma$ is considered to be a characteristic material property. For brittle fracture in effectively 2D materials, the principle of small-scale yielding allows us to concentrate on the singular region surrounding the tip of a crack, so long as all dissipation is contained within a small scale encompassed within the singular region. When rate-dependent dissipation is involved in crack propagation, $\Gamma$ will be dependent on the crack speed, $v$. In this sense, energy balance is generalized to all crack speeds, $\Gamma (v) = G(v)$. If $\Gamma (v)$ is known and $G(v)$ can be calculated as a function of $v$, one can predict the motion of the crack tip. Once the crack motion ensues, the dynamics of simple cracks are entirely described by energy balance; \citet{goldman2010acquisition} showed that LEFM provides an excellent quantitative description of the motion of a crack tip under conditions of either a semi-infinite crack propagating in an infinite medium or an infinitely long strip. Moreover, the rupture of a frictional interface (or earthquake dynamics) is described in both form and motion \citep{svetlizky2014classical,svetlizsky2017eom} by the classical singular solutions for mode II cracks.
 
There are several ways to calculate the $G(v)$. For simple cracks in linear elastic materials, the measurement of crack tip opening displacement (CTOD) can be easily used to calculate $K$ which, via LEFM, yields the value of $G$. In the close vicinity of the crack tip, this calculation should be supplemented by corrections that account for nonlinear elasticity \citep{livne2010near, bouchbinder2014dynamics}. Sufficiently far from the crack tip, the well-known $J$-integral \citep{rice1968path, freund1998dynamic} will also quantitatively provide $G$ by computing the instantaneous rate of energy flow towards the crack tip (in 2D media) through a contour $C$ surrounding the crack tip. $J$ is path-independent in the case of quasi-static and steady-state crack propagation. The $J$-integral can be extended to the case of crack propagation in 3D materials, where the integral becomes domain-independent with $C$ changing to a cylindrical volume around a certain part of a crack front \citep{eriksson2002domain}. In an infinite strip geometry, the translational invariance of the crack in the (steady-state) propagation direction can be utilized to provide a measure of $G$ that is independent of the form of the fields and/or dissipative processes \citep{goldman2010acquisition}. $G$ can also be calculated by considering the crack as a singular defect and $G$ as a configurational force acting on the crack \citep{eshelby1951force, adda1999generalized}. Using this, \citet{adda1999dynamic} suggest a generalized energy (force) balance; balancing $G$ and dissipative forces during crack motion.

In the case of a crack front involving local 3D secondary structure, the local application of the Griffith criterion on the crack front, $G(v, z) = \Gamma(v(z))$, where $z$ represents a spatial point on the crack front, has been widely used in theoretical work to predict the local front motion and stability. For example, \citet{ramanathan1997dynamics} and \citet{morrissey1998crack} studied the interactions of dynamic crack fronts with localized perturbations to the fracture energy using local energy balance. This work predicted a propagating mode within crack fronts, coined `crack front waves', which were later observed  experimentally \citep{sharon2001propagating}. \citet{leblond2019configurational} and \citet{vasudevan2020configurational} combined the Griffith criterion and the principle of local symmetry \citep{goldstein_brittle_1974}, through a heuristic hypothesis of dependence of the fracture energy on the mode mixity ratio, to study the generation of faceted cracks under mode~I + III loading. This work predicts both a low (but finite) threshold for the formation of steps and step drift in the presence of a mode II component. 

$\Gamma (v)$ is considered to be a material-dependent parameter, however it has rarely, if at all, been {\it directly} determined (or measured). Instead, energy balance for simple cracks is {\it used} to determine $\Gamma (v)$, since $G(v)$ can be either calculated or directly measured. In effectively 2D materials, experiments have shown that different methods used to measure $G(v)$ yielded the same result \citep{goldman2010acquisition,sharon1999confirming,scheibert2010brittle}, so the $\Gamma (v)$ that is determined in this way indeed appears to be robust. Can one, however, use (or `trust') analogous measurements of $\Gamma$ for cases where cracks are {\it not} simple? Whereas measurements of $G(v)$ can be performed that are not affected by the nature of a crack front, can a characteristic and solely material-dependent value of $\Gamma$ of a 3D crack (which is {\it independent} of the state of the crack front) be determined via energy balance? A related question is whether energy balance is a local condition (i.e. $G(v,z) =\Gamma(v(z))$. To our knowledge, quantitative measurements of the local fracture energy showing how the local secondary structure of the crack front contributes to $\Gamma$ have not yet been performed. 

In this work, we will address a number of the issues stated above. We will study crack propagation, in polyacrylamide hydrogels, at very low crack speeds, where dynamic effects are negligible. By carefully controlling the crack initiation conditions, we will first show that simple cracks in these gels are universally stable at speeds varying from about 0 to 0.2$c_{R}$. If stringent control is not exercised and slight mixed-mode~I+III perturbations are applied, slow cracks will become segmented and form propagating steps along the fracture front over the same speed range. The co-existence of the single crack and the faceted crack states, therefore, reveals bistability of a crack's state. We then utilize the simple cracks generated to measure, for the first time in these materials, $\Gamma (v)$ at these very low crack speeds, using either the CTOD or the $J$-integral measurements. 

When a crack front develops steps, we will demonstrate that the 3D structure of the crack front significantly increases dissipation and results in an increase of the `apparent' fracture energy that we would assume, were the system entirely 2D. Not only do crack fronts form cusp-like shapes at step locations \citep{kolvin2018topological}, but the dynamic behavior of the entire crack front is affected; under constant $G$ conditions both the mean crack front speeds and lengths continuously change with the step evolution. We find that $G$ is indeed balanced by the total fracture energy, but this only becomes clear when we correctly account for {\it all} of the variations in geometry and dynamics of the crack front that are induced by the steps. 

\section{Materials and Experiments}

\subsection{Properties of the polyacrylamide gels}

Our fracture experiments were performed using polyacrylamide gels, which obey the neo-Hookean elastic constitutive law. The materials are homogeneous, transparent, and incompressible. Crack dynamics in these materials are representative of the broad class of materials that undergo brittle failure \citep{livne2010near, goldman2010acquisition, bouchbinder2014dynamics}. The near-tip fields of propagating cracks are singular and the features characterizing their dynamics (e.g. microbranches,  front waves, equations of motion) are identical to those of other brittle materials \citep{livne2005universality}. Hence, polyacrylamide gels have been used to verify LFEM predictions \citep{livne2010near, goldman2010acquisition} and to investigate the effects of nonlinear elasticity \citep{bouchbinder2014dynamics}. A significant advantage of using these materials to study fracture is that they provide a means to perform direct and precise measurements of the near-tip structure of the fields driving rapid cracks, by slowing crack propagation speeds by nearly three orders of magnitude ($c_R$ in these gels is, for example, 500 times below that of soda-lime glass).

The gels used in this work have a composition of 13.8$\%$ (w/v) acrylamide/bis-acrylamide with a 2.6$\%$ (w/v) cross-linker concentration, providing a Young's modulus $E$ = 105.6 $\pm$ 4.2 kPa and shear wave speed $c_{s}$ = 5.9 $\pm$ 0.15 m/s. The Poisson ratio of 0.5 yields a plane stress Rayleigh wave speed, $c_{R}$, of 5.5 $\pm$ 0.15 m/s. This is the same gel composition used in much previous work on brittle fracture \citep{livne2005universality, livne2010near, goldman2010acquisition}. Our samples are of long-strip geometries with typical dimensions $L_{0} \times b \times w$ of 40 $\times$ 20 $\times$ 1 mm along the crack propagation $x$, tensile loading $y$, and sample thickness directions $z$, respectively (see Fig. \ref{fig1.fig}). A pre-crack of 5 $\pm$ 1 mm along the $x$ direction is introduced at one of the edges of each sample, midway between its vertical ($y$) boundaries. 

\subsection{Preparing the initial `pre-crack'}
\label{precrack}

The form of the imposed pre-cracks significantly affects a crack's initial propagation mode. To generate pure mode~I propagation, special efforts were required to create nearly pure mode~I pre-cracks, whose entire fracture plane is, as closely as possible, within a single $xz$ plane aligned normal to the loading ($y$) direction. These `clean' pre-cracks were formed by forcing initial cracks to arrest while imposing external guiding of the initial crack direction. To create clean pre-cracks, we first adhered a thin layer of PDMS to one of the gel sample faces to act as a `guide'. This guide had the same thickness as the gel sample that was adhered to it and contained a cut with the desired extension length of the pre-crack. A shorter pre-crack along the $x$ direction was created within the gel, by the application of a scalpel having its $xz$ plane located at the $y$ location of the cut in the guide. The $xz$ plane formed by the scalpel needed to be both oriented correctly and as mirror-like as possible so that a pure mode I crack could be generated after a short extension. The composite gels/PDMS sheet was then non-uniformly stretched under tension, with the maximal stretch at the position of the cut. Since the shorter pre-crack within the  gel was free of adhesive constraints and the PDMS layer is much tougher than the gels, only the extension of the pre-crack within the gel sample was triggered. This pre-crack then extended until encountering the notch tip determined by the PDMS guide. Beyond this point, the PDMS layer constrained the opening displacement of the crack and, consequently, arrested the crack. The PDMS guide both determined the length of the pre-crack, and, importantly, forced it to be constrained within the desired initial plane. Once the initial crack was formed, the applied tension was reduced to zero and the PDMS layer was removed.

To generate faceted cracks, the PDMS guide was simply not used. This caused pre-cracks to be slightly tilted in planes {\it not} normal to the $y$ direction. Any tilt produced a small local mode~III component at the tip of the initial crack \citep{ronsin2014crack} that was sufficient to excite facets. We also found that mode~II components that were externally imposed onto a clean initial crack face would {\it not} excite facets.

Once the pre-crack was formed,  fracture was initiated in mode I by a slow and uniform displacement of the vertical boundaries until reaching the fracture threshold, whose value was dictated by the length of the pre-crack. In this way, experiments were controllably performed for a range of imposed strains. In the long strip configuration, cracks accelerate at the early stage and reach nearly steady-state propagation after communicating with the sample boundaries \citep{goldman2010acquisition}.

\subsection{Fracture experiments surrounded by air}

\begin{figure*}[]
\centering
\includegraphics[width=1\linewidth]{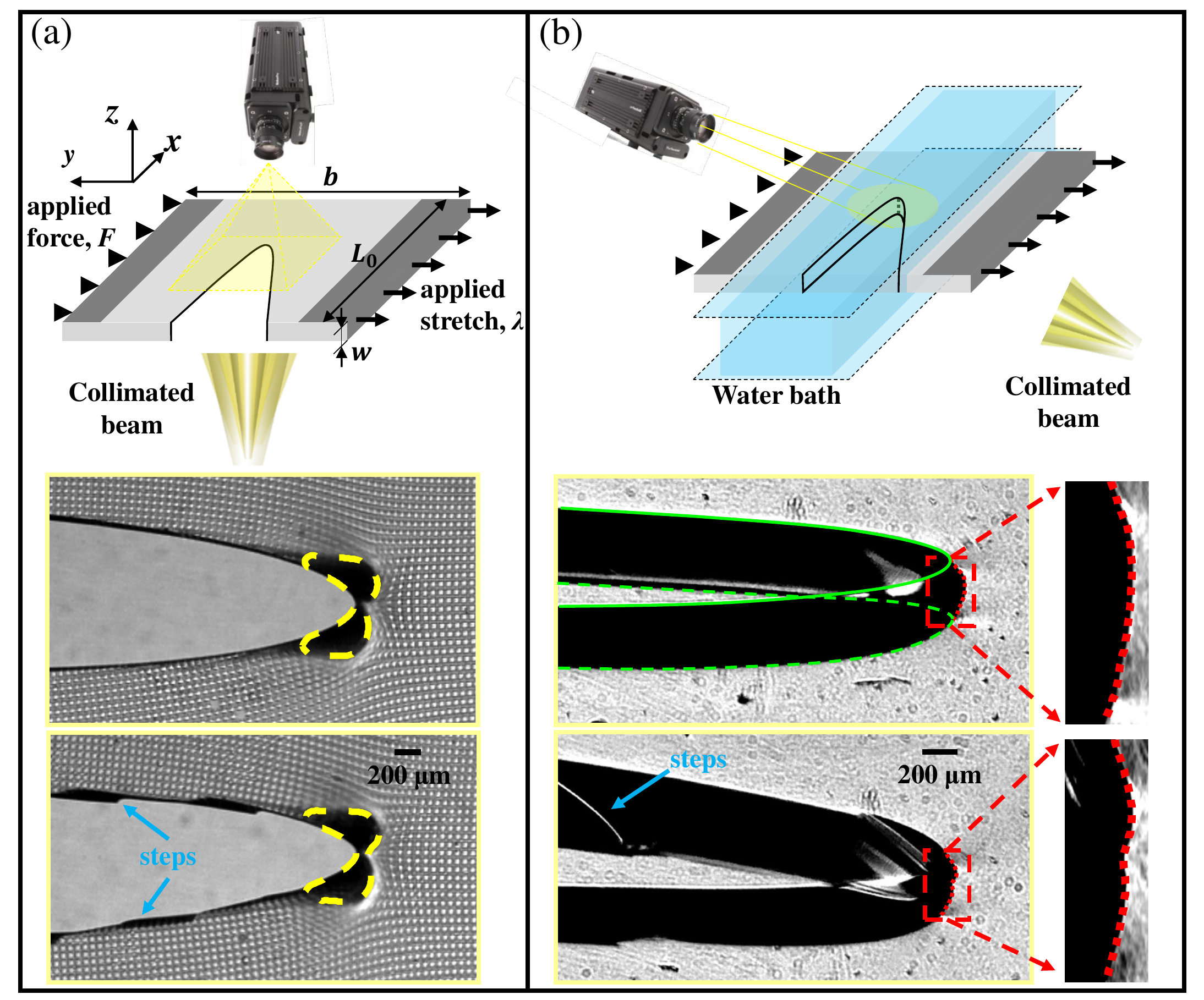}
\caption{\textbf{Experimental setup.} Experiments were performed with transparent polyacrylamide gel sheets under quasi-static tensile loading. The gel sheets were surrounded by either air or water layers. (a), In the air, gels were illuminated via a collimated beam of incoherent light that was transmitted normal to sheet surface (from the bottom face) to produce a shadowgraphy image of  both the crack opening and a grid that was imprinted on one surface of the gel ($xy$ plane). Dynamic cracks were initiated from a pre-crack located at the center of one of the sample edges. The shadowgraph images formed by the fracture process were captured with a high-speed camera mounted above the sample. Lower panels: Example of a pure mode I simple crack (top) and a faceted crack (bottom) imaged from above. Two of the steps formed on the fracture surface of the bottom image are noted. The caustics in the vicinity of the crack tip (highlighted by the yellow dashed curves) are due to a lensing effect; the high stresses near the crack tip cause the gel to contract in the $z$ direction. (b), Some experiments were performed in a water bath to eliminate this lensing effect and the resultant caustic. The lighting and camera were mounted at 45$^{\circ}$ to the $yz$ plane. This enabled the entire crack front to be visualized by shadowgraphy, together with the CTOD. Bottom panels: a simple mode I crack front is smooth (top), while the front locally forms a cusp-like shape (bottom) surrounding any steps formed. In the upper panel, the crack tip opening displacements of the upper and lower surfaces of the simple crack are highlighted by the green full and dashed lines, respectively. Both crack fronts are denoted by red dotted lines. The black sections behind the crack front correspond to the planar upper and lower fracture surfaces where the transmitted light was refracted away from the camera. Steps on the fracture surface (blue arrow) are observable due to light that is scattered into the camera by step edges.}
\label{fig1.fig}
\end{figure*}

Crack motion together with the surrounding displacement fields were measured using a fast camera (IDT-Y7) having a spatial resolution of 1920 $\times$ 1080 pixels and a frame rate of 8000 Hz. The camera was mounted above the sample and normal to its plane and imaged an area 10.7 $\times$ 6 mm$^{2}$ that initiated a few mm's beyond the pre-crack. The imaged area was illuminated via a collimated light beam directed normal to the sample plane from below. 

Around the crack tip, large deformation gradients along the $z$ direction appear within the singular region. These gradients are due to strong material contraction in $z$, caused by gel incompressibility,  that must balance the large extensions in the $xy$ plane. This strong contraction gives rise to lensing effects; light, which is strongly refracted in the near-tip singular region, does not reach the camera. This creates the caustics within the image, the black regions surrounding the tip that are highlighted by the yellow dashed curves in the bottom panels of Fig.~\ref{fig1.fig}a. In the past, these caustics have provided an optical tool that was utilized to study the singularities of the stress field around the crack tip \citep{Manogg1964, theocaris1970local}. Here, we used the centroid of these caustics to both determine the crack tip location and calculate the crack speed. This method is consistent with the use of the tip of the parabolic crack opening for the single crack and provides improved precision for crack speed measurements of faceted cracks, as the caustic centroid provides the instantaneous mean position of the crack front in $z$. In the following sections, as we will be describing crack fronts, to avoid confusion we define $v$ as the mean crack front speed in $z$ and $v(z)$ as the normal velocity to the crack front at each spatial location, $z$.

We measure the displacement field around the crack by imprinting on one surface of the gel sample (see \citet{boue2015failing}), a shallow square grid (depth 2 $\mu$m) having a lattice spacing of 60$\mu$m. This was accomplished as follows. We cast the gels in a  mold formed by two glass plates separated by a (typically 1mm) spacer. On the $xy$ surface of one of these plates, we embossed a rectangular grid formed by lithographical printing of a spin-coated epoxy layer. Upon casting, this grid mesh was imprinted on one gel surface.  When a crack propagated across the measurement area, each frame of the camera captured the instantaneous image (through shadowgraphy) of the distorted grid. The location of each grid point in the deformed field was determined by its center with a resolution of $\sim$1 $\mu$m. The deformation field surrounding the crack was obtained by comparing the position of the grid points in the deformed frame to their position in the reference (deformation-free) frame.
Examples of propagating simple and faceted cracks with their respective deformed grid patterns are presented in Fig.~\ref{fig1.fig}a.

\subsection{Fracture experiments surrounded by water}

To follow the crack front dynamics along the thickness ($z$) direction, while, in parallel, measuring the mean location of the crack front in the $xy$ plane, we developed a slightly different optical technique. To this end, we needed to both remove the caustics as well as enable optical access to the crack front during propagation. Gel samples without grids were used. The transparency of the gels provided the possibility to observe the whole crack front, when oblique imaging is used. When the samples are bounded by air, however, the crack front is hidden by the caustics formed in the vicinity of the crack front. Since the gels consist of cross-linked polymers immersed in water, their measured refractive index (1.365) is nearly perfectly matched to that of the water (1.333). Hence, as illustrated in Fig.~\ref{fig1.fig}b, we were able to eliminate the appearance of caustics during fracture experiments by surrounding the sample with a water bath. We note that variations of the fracture energy due to the surface tension (72.8mN/m) of the surrounding water are $<1\%$ and therefore negligible.

 The crack front motion was measured by the fast camera using highly magnified images (a field of 6.1 $\times$ 3.5 mm$^{2}$ was mapped to the camera's 1920 $\times$ 1080 pixel resolution) with a frame rate of 7000 Hz. As presented in Fig.~\ref{fig1.fig}b,  the dynamics of the whole crack front could be captured by mounting both the camera and collimated beam at an angle of 45$^{\circ}$ relative to the $xy$ plane. 

Fig.~\ref{fig1.fig}b presents snapshots of both a single crack and a faceted crack developing steps. The CTOD of the top and bottom surfaces of the single crack are highlighted by the green full and dashed lines, respectively.  As the light passing through the crack opening surface is refracted away from the camera, shadowgraphy could be used to image the crack front. Owing to the small mismatch of the refractive index between the gels and water, slight caustics can be observed at the two extremes of the crack front. These permit us to easily determine the crack front boundaries (see edges of dotted lines in the bottom panels of \ref{fig1.fig}b). Both the local crack front velocity and the mean crack front speed could be determined by using the instantaneous crack front shapes and positions.

When a crack forms steps, their characteristic cusp-like shapes within the crack front can be observed, as reported by \citet{kolvin2018topological}. In addition, in each frame, we are able to observe the step edge left behind the front, as highlighted in Fig.~\ref{fig1.fig}b (bottom inset). This is possible because some of the transmitted light is scattered by the step edge into the camera. To characterize the topography of the steps, immediately following experiments where a faceted crack was formed, we created a cast of the fracture surface using polyvinyl siloxane. These casts are able to reproduce the surface topography at microscopic levels. We then measured the fracture surface casts using an optical profilometer with an in-plane resolution of 2 $\mu$m and out-of-plane resolution of $\sim$0.1 $\mu$m. 

\section{Results}

\subsection{Bistability of simple and faceted cracks}

\begin{figure*}[]
\centering
\includegraphics[width=1\linewidth]{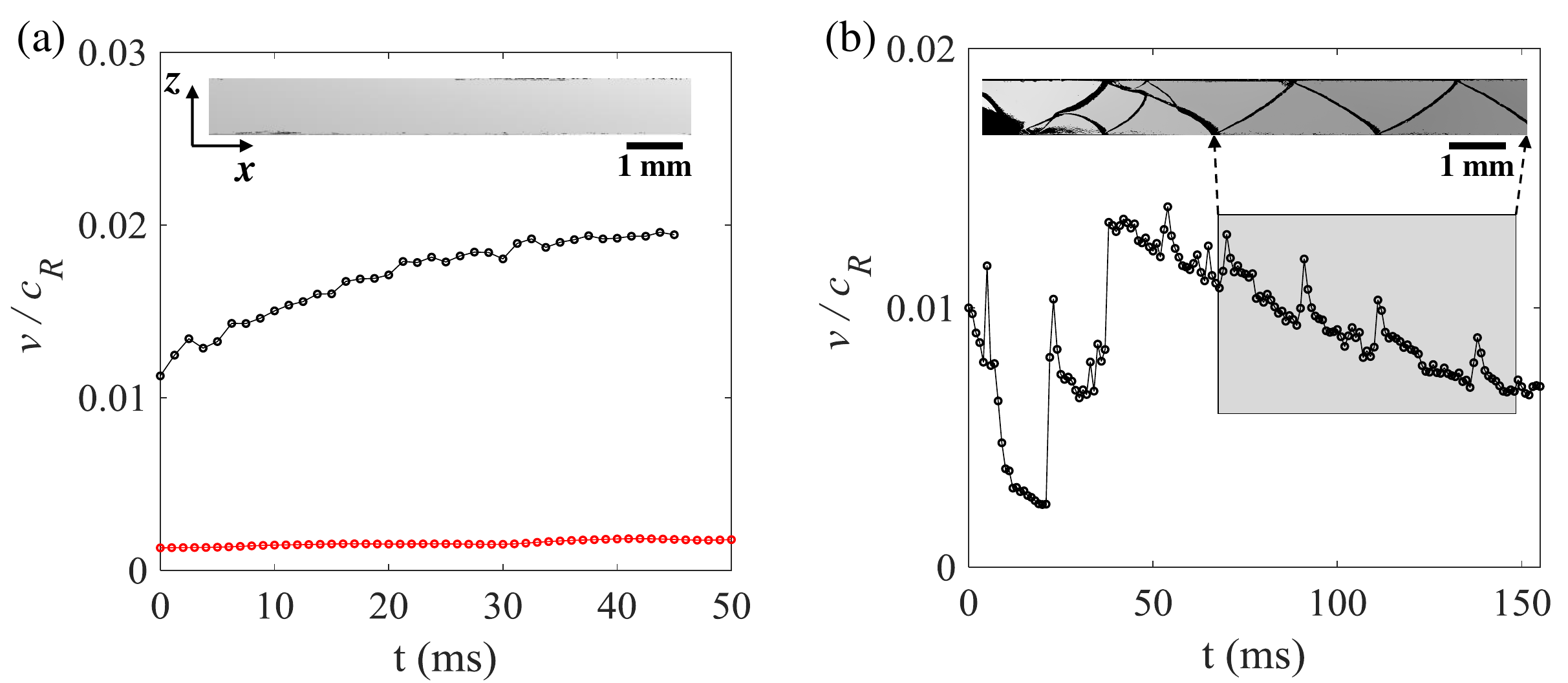}
\caption{\textbf{Bistability of slow cracks.} Typical crack speeds as a function of time for simple (a) and faceted (b) cracks. (a), A simple crack, triggered from a clean pre-crack located along the $xz$ plane under pure mode~I loading conditions, generates a mirror-like fracture surface (inset). Presented are the dynamics of two typical simple cracks that were driven by imposed stretches of 1.07 (black line) and 1.063 (red line). The former (black line) slowly accelerated before reaching steady-state propagation of $v$=10.9 mm/s = 0.02$c_{R}$. In the latter experiment (red line) the crack propagated at a speed of $v \sim$1.0 mm/s = 0.002$c_{R}$. (b), A faceted crack was triggered via a pre-crack that was slightly tilted away from the $xz$ plane, and propagated under applied tensile loading condition with an imposed stretch of 1.068. The tilted pre-crack generated a mixed-mode (I+III) initial condition near its tip. This experiment formed a faceted fracture surface (inset). In contrast to the smooth dynamics of the simple cracks in (a), the mean (in $z$) crack front dynamics were erratic, reflecting the complex dynamics of the initial steps and, later, of a single step (shaded region) that propagated within the crack front. }
\label{fig2.fig}
\end{figure*}

In these materials, `simple' mirror-like cracks in mode~I have never, to our knowledge, been observed for low velocities. As explained in Section~\ref{precrack}, we were able to achieve simple crack states for low velocities from 0-0.2$c_R$ by very carefully setting the initial conditions of the pre-crack prior to application of stresses. Fig.~\ref{fig2.fig}a presents an example of the crack speed, $v$, as a function of time for two typical simple cracks, that propagated at steady-state velocities of $v=0.002c_R$ and $0.02c_R$. Stable simple crack propagation generates a mirror-like fracture surface, as shown in the inset of Fig.~\ref{fig2.fig}a. We find that, once simple cracks are excited, they remain `simple' for any velocity up until the formation of micro-branches. This implies that pure mode~I cracks in gels can exist at any slow crack speed.

Faceted crack propagation at a very low speed was achieved by initiating fracture with slightly tilted pre-cracks, which generated  mixed-mode I + III initial conditions. Fig.~\ref{fig2.fig}b presents an example of crack dynamics during the propagation of a faceted crack. The crack speed is highly fluctuating and the fluctuations are correlated with the presence of crack segmentation. The segmentation of the fracture surface is the result of step formation within the crack front. The crack develops out-of-plane steps, which propagate along the crack front at an angle of about 43$^{\circ}$ relative to the local front normal, as reported by \citet{kolvin2018topological}. The traces of traveling steps form step-lines on the fracture surface. In general, a crack will develop multiple steps immediately after initiation, when subjected to mixed mode I+III perturbations. As noted previously \citep{ronsin2014crack, pham2017formation, pham2016growth}, initial steps have complex behavior (as seen in, e.g., Fig.~\ref{fig2.fig}b). Steps may separate, coarsen and/or disappear upon interaction. When steps encounter a free surface, they are often reflected; steps approaching a free surface will change direction and propagate to the other free boundary. Such repeated step reflection creates a periodic step-line on the fracture surface (Fig.~\ref{fig2.fig}b). In parallel, the mean crack speed along the sample width ($z$) oscillates in phase with step reflections (see the shaded region of Fig.~\ref{fig2.fig}b). 

Fig.~\ref{fig2.fig} also demonstrates crack {\it bistability} at low speeds. Both simple and faceted cracks propagate within the same range of velocities and applied loads. When initiated by mixed mode initial states, faceted crack states may appear from speeds of nearly zero. Faceted crack states will generally disappear when crack speeds increase to sufficiently high values. Without taking special care in forming pre-cracks, mirror-like cracks will often appear when a crack jumps to $0.1-0.2c_R$ upon initiation. Empirically, cracks in polyacrylamide gels appear to be immune to the precise nature of the initial pre-crack when they jump to this velocity range \citep{livne2005universality,goldman2010acquisition}. Faceted cracks could also transition directly to micro-branches \citep{kolvin2017nonlinear} in this velocity range. At much higher velocities ($0.2-0.95c_R$) \citep{livne2005universality} bistability between simple cracks and micro-branches may also take place, but faceted cracks are not observed.

\subsection{Simple cracks: energy flux and fracture toughness}

Let us first focus on the simple crack state. Crack propagation is understood to be governed by energy balance. The crack speed, $v$, is governed by balancing the energy release rate, $G$, into the crack front with the fracture energy $\Gamma (v)$, which characterizes the velocity-dependent energy dissipation of the crack. The energy consumption within the dissipative zone (per sample thickness) can be evaluated in a number of ways. Owing to the nearly steady-state crack propagation at the very low speeds, we compute the energy flux into any closed contour, $C$, surrounding the crack tip using the $J$-integral \citep{rice1968mathematical}:
\begin{equation}
J = \int_{C} \left[U (\textbf{F})n_{x} - \sigma_{ij}n_{j}\frac{\partial u_{i}}{\partial x}\right] dC\;,
\label{eq-J.eq}
\end{equation}
where $U (\textbf{F})$ is the strain energy density, $n_i$ stands for the components of outer normal vector of $C$, $u_i$ and $\sigma_{ij}$ are the 2D displacement and stress field components. In our experiments, the gels were deformed under, effectively, plane stress conditions. This yields a neo-Hookean strain energy density, $U (\textbf{F}) = \frac{\mu}{2}[tr(\textbf{F}^{T}\textbf{F}) + (det \textbf{F})^{-2} - 3]$, where \textbf{F} is the 2D deformation gradient tensor, $F_{ij} = \delta_{ij} +\partial u_{i}/\partial x_j$. The calculation of $J$ is path-independent so long as the contour encircles the entire dissipative region. This condition also ensures $G \equiv J$, that is the energy release rate $G$, a local quantity of the crack front, is given by the $J$-integral, a far field quantity.

\begin{figure*}[]
\centering
\includegraphics[width=1\linewidth]{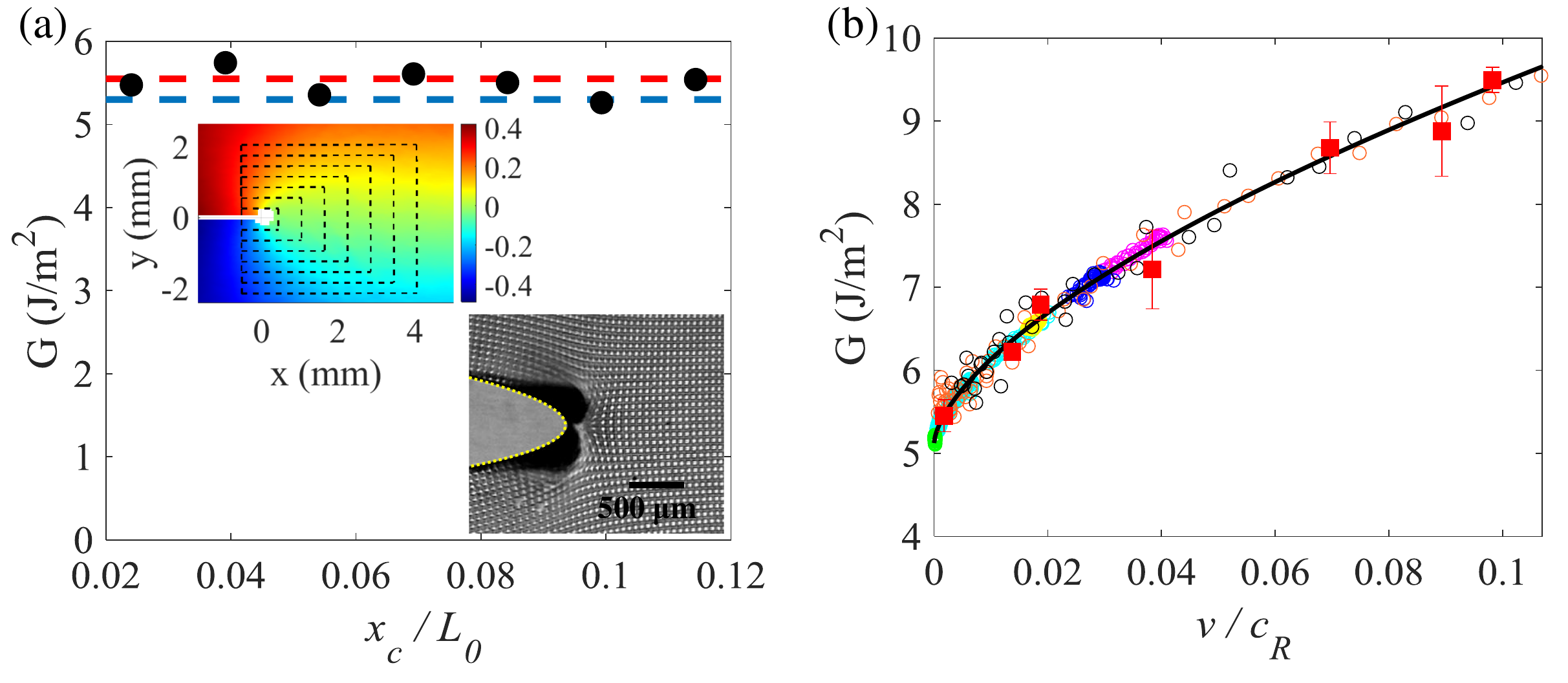}
\caption{\textbf{Measurements of the energy flux of simple crack states.} (a), Calculation of the energy flux $G$ for a simple crack propagating at a speed of 1.0 mm/s = 0.002$c_{R}$. This sample was loaded under mode~I conditions, with an imposed stretch of 1.063. $G$, is computed using both the $J$-integral over different contours (black dashed lines in upper inset) and the CTOD (yellow dashed line in lower inset). Here, the $x$ axis is the spatial extension of each contour $C$ in the upper inset along the crack propagation direction normalized by $L_{0}$ (the color map corresponds to $u_{yy}$), while the black dots represent the value of $G$ calculated via the $J$-integral for each contour. The red dashed line is the value of $G$ calculated using the CTOD presented in the lower inset. Both measurements are within 4\% of the value of 5.28 J/m$^{2}$ that corresponds to the total work measured directly from the force-displacement loading curve (blue dashed line). (b) $G$ as a function of the crack speed, $v$, for multiple experiments. $G$ calculated using CTOD (circles) and $J$-integral (squares) are in perfect agreement. Colors correspond to different experiments. The solid line is a guide to the eye and corresponds to a spline fitting of the data.}
\label{fig3.fig}
\end{figure*} 

Using the displacement field measured with the grid mesh, $G$ could be computed by means of Eq.~(\ref{eq-J.eq}). An example of the calculation of $G$ (through different contours surrounding the crack tip) for a simple crack propagating at $v = 0.002c_{R}$ is presented in Fig.~\ref{fig3.fig}a. $G$ is, indeed, seen to be path-independent, even though the smallest enclosed area is below a few hundred $\mu$m$^{2}$. This is consistent with \citet{livne2010near}, who showed the dissipative scale to be within $\sim$20 $\mu$m. It's worth noting that, as the $J$-integral is measured along the free surface, it represents the energy flux per unit sample thickness. The path-independence of $J$ reveals that there are no noticeable 3D effects and no plastic or extra dissipative effects at the smallest measured scale.

We can use the crack tip opening displacement (CTOD) to validate our measurements of $G$. LEFM predicts that the opening displacement of a mode I crack tip can be described by a parabolic shape, for scales beyond the nonlinear elastic region adjacent to the crack tip \citep{livne2010near}. An example is presented in the Fig.~\ref{fig3.fig}a, where the excellent parabolic fit of the CTOD implies that, at these very low velocities, the size of the nonlinear region is below $\sim$30 $\mu$m. LEFM relates the curvature $a$ of the mode I crack tip to the stress intensity factor $K_{I}$ through 
\begin{equation}
a = \frac{32\pi\mu^{2}(1 + T / 3\mu)}{[K_{I}\Omega_{y}(\theta = \pi; v)]^{2}}\;,
\label{eq-CTOD.eq}
\end{equation}
where the moving coordinates $(r, \theta)$ are centered at the crack tip ($\theta$ = 0 is the crack propagation direction), and $\Omega_{y}(\theta; v)$ is a universal function of $\theta$ and $v$ \citep{freund1998dynamic}. $T$ in Eq.~(\ref{eq-CTOD.eq}) is the `$T$ stress', which was calculated for the strip configuration by \citet{katzav2007theory}. In Eq.~(\ref{eq-CTOD.eq}) we ignored the background strain dependence of the CTOD, which gives a correction within 5$\%$ for the low crack speeds in these experiments \citep{boue2015failing}. Using the measured $a$, Eq.~(\ref{eq-CTOD.eq}) provides $K_{I}$. For plane stress conditions, $G$ is then given by:
\begin{equation}
G(v) = \frac{1}{E}A(v) K_{I}^{2}(v)\;,
\label{eq-Gv.eq}
\end{equation}
where $A(v)$ is a known universal function of $v$ satisfying $A(0)= 1$ \citep{freund1998dynamic}. Since the CTOD is obtained from the average projection of the crack opening through the sample thickness, the measured $G$ also corresponds to the effective 2D energy flux. The value of $G$ derived from Eq.~\ref{eq-Gv.eq} (red dashed line in Fig.~\ref{fig3.fig}a) indeed agrees well with the independent measurements using the $J$-integral. The value of $G$ is further validated using the force-displacement loading curve (see blue dashed line in Fig. 3a) through, $G = \int_{1}^{\lambda_{max}} \sigma(\lambda) d\lambda \cdot b$, where $\sigma$ is the nominal applied stress given by $F/wL$. Measurements of $G$, using both the $J$-integral and CTOD, for different cracks with the crack speed varying from about 0 to 0.1$c_{R}$ are shown in Fig.~\ref{fig3.fig}b. 

\begin{figure*}[]
\centering
\includegraphics[width=1\linewidth]{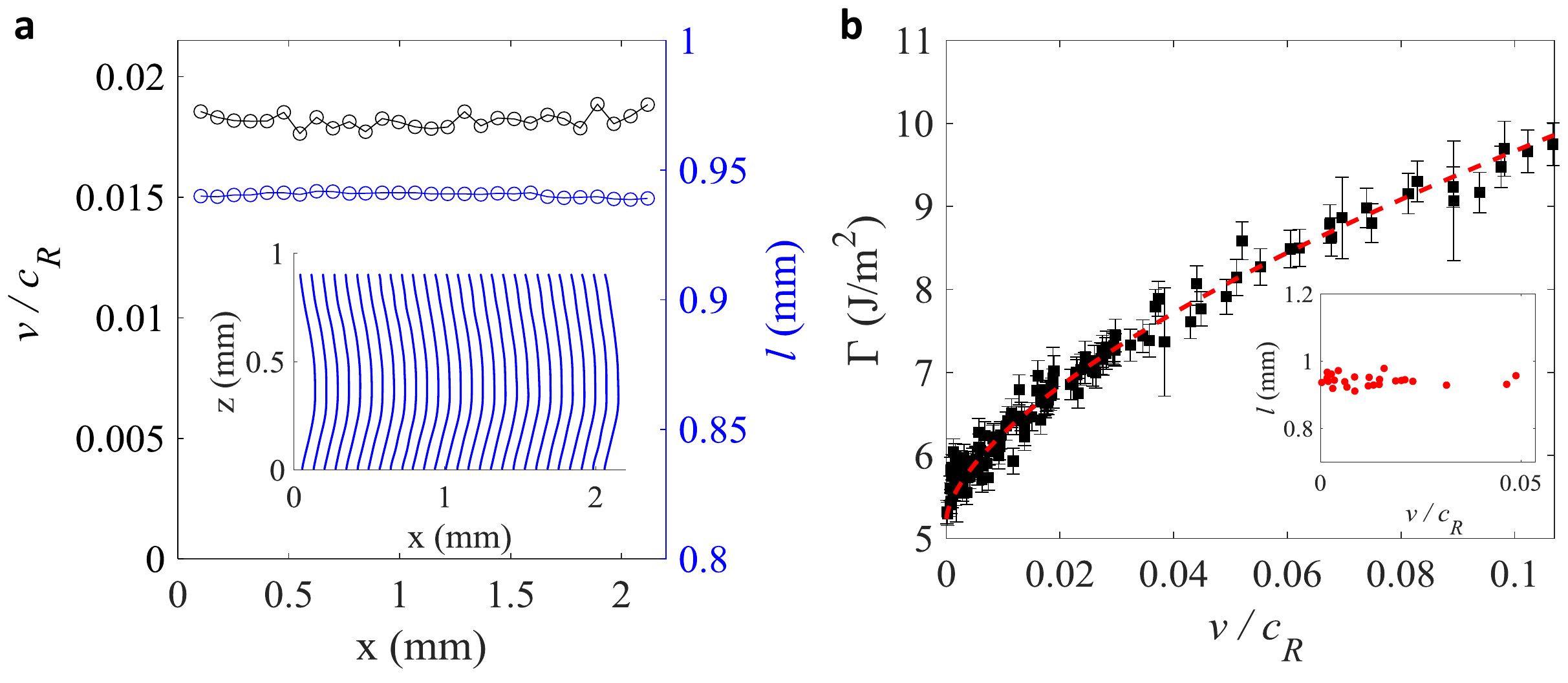}
\caption{\textbf{Dependence of the fracture energy on the crack speed.} (a), Crack speed (black symbols) and crack front length (blue symbols) as a function of a crack's position along the $x$ direction of a typical simple crack. Inset: Successive crack front shapes at different positions separated by time intervals of 0.714 ms. Note that, due to Poisson contraction, the measured sample thickness, $w$, is below the zero strain sample width of 1 mm. The crack front curvature typically increases the crack front length, $l$, relative to a nominally straight crack front, by about 5.5$\%$. (b), Measured fracture energy $\Gamma (v)$ as a function of the crack speed, $v$. The fracture energy $\Gamma=G\cdot w/l$ differs from the bold line in Fig.~\ref{fig3.fig}b as it takes into account the crack front curvature. The uncertainty in $\Gamma$ is due to the uncertainty in $l$. The red dashed line corresponds to a spline fitting of the data and is a guide to the eye. Inset: Crack front lengths $l$ of simple cracks with $v$ for imposed stretches between $1.06<\lambda < 1.09$.}
\label{fig4.fig}
\end{figure*}

Fig.~\ref{fig4.fig}a presents measurements of a typical simple crack front propagating at a steady-state velocity of $v = 0.018c_{R}$, as determined by the mean front position in $z$. The corresponding sequence of instantaneous crack fronts shows that simple crack fronts possess invariant shapes whose lengths $l$ are constant. We note that i), due to the incompressibility of the materials, the sample widths in the deformed (lab) frame, $w$, contract (via Poisson contraction) in the $z$ dimension. ii), Crack fronts of simple cracks are not straight, but curved. This curvature typically increases the total front length by about 5.5$\%$. Since the variation of the imposed stretches, $\lambda$, is small, crack front lengths in our experiments are nearly constant at about 0.945 $\pm$ 0.03 mm (inset of Fig.~\ref{fig4.fig}b) over the  range of measured $v$. Over this range, both $l$  and simple crack front shapes are invariant. The curved crack front coupled with the sample width contraction implies that the measurements of $G(v)$ (as shown in Fig.~\ref{fig3.fig}b) do not provide $\Gamma(v)$ unless the crack front length is properly accounted for. $\Gamma(v)$ and $G$ are related via the total energy balance of the integral crack: 
\begin{equation}
\int_{w} G(v, z) dz = \int_{l} \Gamma(v(z)) dz\;,
\label{eq-G_Gamma.eq}
\end{equation}
where $G(v, z)$ is the energy flux into the crack front per unit sample thickness measured using Eq.~(\ref{eq-J.eq}) with $C$ far from the crack. Since, far from the crack, $G(v)$ is independent of $z$ and the invariance of the front shape implies that the local crack speed, $v(z)$ equals the mean crack front speed $v$, Eq.~(\ref{eq-G_Gamma.eq}) can be written as: 
\begin{equation}
G(v) w = \Gamma(v) \int_{l}dz\;.
\label{eq-Gw_Gammal.eq}
\end{equation}
Using the measured values of $l$, we derive $\Gamma(v)$ using our measurements of $G(v)$ (Fig.~\ref{fig3.fig}b) and the correction factor $w/l$  as input. In contrast to the fracture energy of glass \citep{sharon1999confirming}, where $\Gamma(v)$ only weakly varies with $v$, Fig.~\ref{fig4.fig}b shows that $\Gamma(v)$ in gels is a strongly rate-dependent function at low speeds. In the extreme low-speed range ($0<v< 0.1c_{R}$), $\Gamma(v)$ is a significantly stronger function of $v$ than for $0.1<v<c_{R}$ \citep{livne2010near, boue2015failing}. Since no crack front structure is observed, we suspect that the rapid increase of $\Gamma(v)$ with $v$ is related to some (as yet, unclear) nonlinear dissipation mechanism involved in breaking the polymer chains that bind the gels. It is interesting that a similarly rapid increase in $\Gamma$ with $v$ has also been observed in other polymers for low fracture velocities, such as PMMA \citep{scheibert2010brittle} and multimaterial 3D-printed polymers \citep{albertini2021effective}.

\subsection{Energy flux and the dynamics of faceted cracks}

\begin{figure*}[]
\centering
\includegraphics[width=1\linewidth]{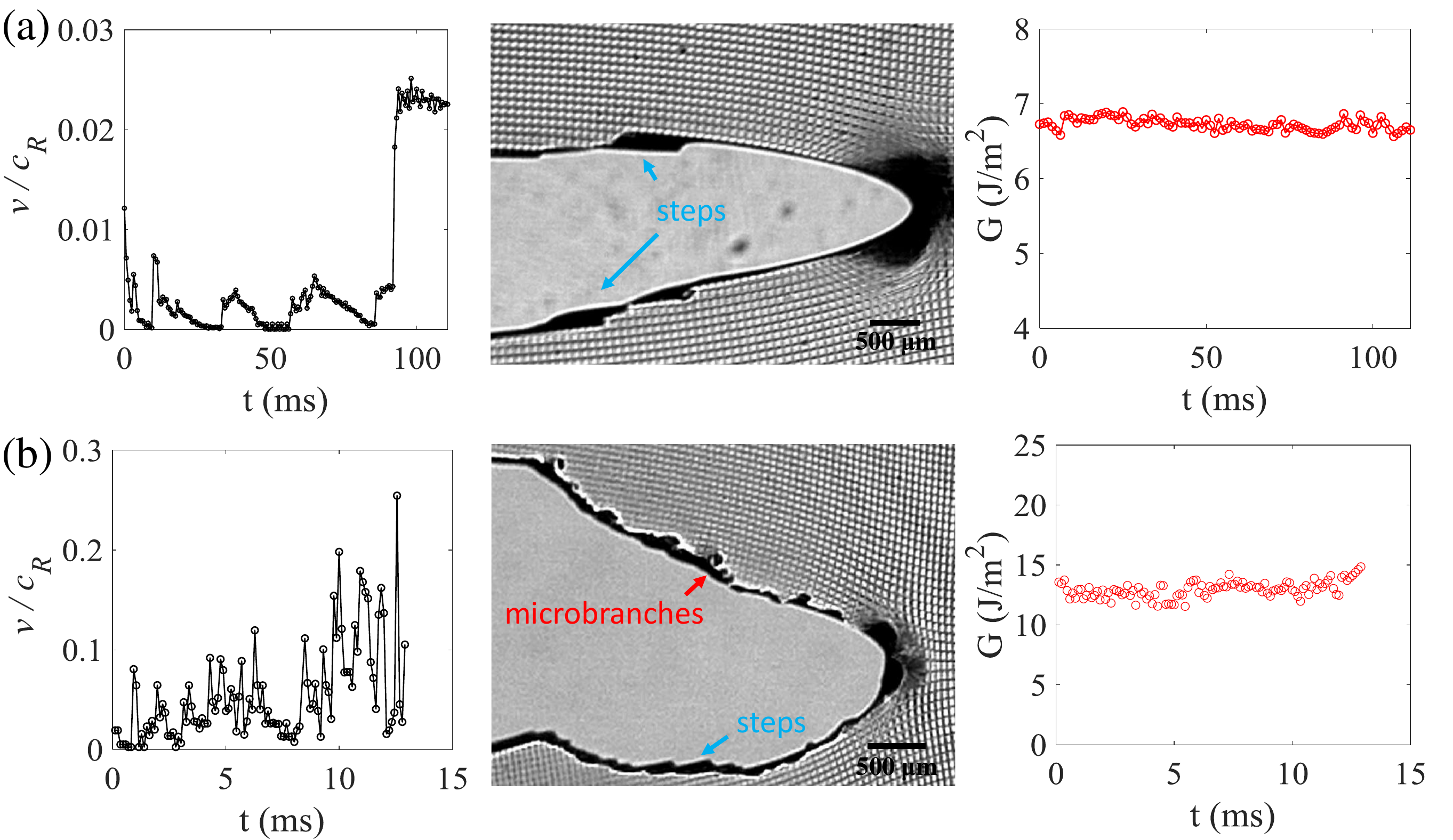}
\caption{\textbf{The energy flux $G$ and dynamics of faceted cracks.} Two examples of faceted crack dynamics, each with a constant value of $G$. Applied uniaxial stretches, (a) $\lambda=1.07$ and (b) $\lambda=1.1$. (a) $v(t)$ of a faceted crack that develops steps upon initiation and undergoes apparent stick-propagate motion before transitioning, at $t=92 ms$, to a simple crack and (b) a faceted crack that transitioned to a micro-branching state, at higher applied strains. Center panels: Snapshots of these cracks in the $xy$ plane for (a) $t=105 ms$ and (b) $t=12 ms$. Typical steps and micro-branches that were generated upon the death of the steps are highlighted. Right panels: $G$, measured via Eq.~(\ref{eq-J.eq}), is constant throughout the cracks' propagation in both examples. 
}
\label{fig5.fig}
\end{figure*}
 
Let us now consider faceted cracks (Fig.~\ref{fig2.fig}b), which are initiated via tilted pre-cracks (see Section. \ref{precrack}) that generate local mixed mode I+III conditions. The propagating steps along the crack front that form the facets locally increase the crack front length, thereby leading to increased local dissipation. \citet{kolvin2018topological} analyzed the in-plane dynamics of steps and showed that the local fracture energy increase caused by steps induces geometric curvature within an, otherwise, approximately straight crack front. Induced front curvature resulting from a spatially implanted step in fracture energy has also been observed in static cracks \citep{chopin2011crack}. Both can be described well by LEFM. We now show that step dynamics change not only the local behavior of the crack front, but the dynamics of the entire crack front. One such example was presented in Fig.~\ref{fig2.fig}b; even a single step propagating through the crack front gives rise to apparently unstable front propagation. 

We first consider the energy flux into faceted crack fronts during non-steady front dynamics.
Fig.~\ref{fig5.fig} presents two measurements of faceted cracks propagating under different stretch levels. In Fig.~\ref{fig5.fig}a a faceted crack (stretch of $\lambda=1.07$) undergoes nearly `stick-propagate' motion (although $v$, though very small is always finite). Here, a single step, which is reflected by each free surface ($z = 0,w$), moves within the crack front. At $t = 92 ms$ the step disappeared and the front became a simple crack. Upon the step's disappearance, the front's propagation velocity instantaneously jumped by a factor of over 5 and continued to propagate as a simple crack in steady state motion. As in Fig.~\ref{fig3.fig}a, we use the $J$-integral to obtain $G$ throughout this unsteady motion. As for simple cracks, measured $J$-integrals are path-independent. We find that throughout this entire complex scenario, $G$ remained constant (Fig.~\ref{fig5.fig}a- right). We note that values of $G$ calculated via the 2D contours are valid, so long as the contours  are sufficiently far from the crack front to enable any fluctuations in $z$ of the strain fields to be negligible. Moreover, the value of $G$ entirely determined which value of $v$ the simple crack acquired after the jump; $v=0.023c_{R}$ corresponded precisely to the $G(v)$ given in Fig.~\ref{fig3.fig}b. 

As the mean propagation velocity increases, the complexity of the motion of faceted cracks generally increases; step reflection is more frequent (as their motion in $z$ scales with $v$) and spontaneous step nucleation or even spontaneous transitions to micro-branching may take place. The resulting fluctuations in $v$ become more frequent and highly erratic. Fig.~\ref{fig5.fig}b presents a particular case of a faceted crack that initiated from a rough pre-crack under a relatively high stretch, $\lambda=1.1$. Very complex motion ensued, which included both step motion and step-generated micro-branches \citep{kolvin2017nonlinear} that broke the up-down symmetry within fracture surfaces that occurs when only steps propagate.  With the generation, transitions, and death of steps, the crack front's motion became so strongly irregular that the crack's overall propagation direction changed. Despite these highly complex dynamics, Fig.~\ref{fig5.fig}b  demonstrates that $G$ remained constant at every instant. Hence, the global dissipation of these highly erratic crack fronts is invariant, even though large variations and re-distributions of {\it local} energy dissipation frequently took place. 

Since steps have an inherently 3D structure, one may ask whether 2D measurements, as described by Eq.~(\ref{eq-J.eq}), correctly evaluate the energy flux to a highly complex 3D system. In light of the examples presented in Fig.~\ref{fig5.fig}, it is puzzling why both the geometry and dynamics of a step-forming crack front are so rapidly changing, despite the fact that the global value of $G$ does not vary at all. How does the crack front adapt to maintain invariant global energy dissipation? How does the local structure within the crack front contribute to the energy dissipation? To address these questions, we now examine the local behavior of a step-forming crack front using the 3D measurement configuration.

\subsection{Dissipative mechanisms of step-forming cracks}

\begin{figure*}[]
\centering
\includegraphics[width=1\linewidth]{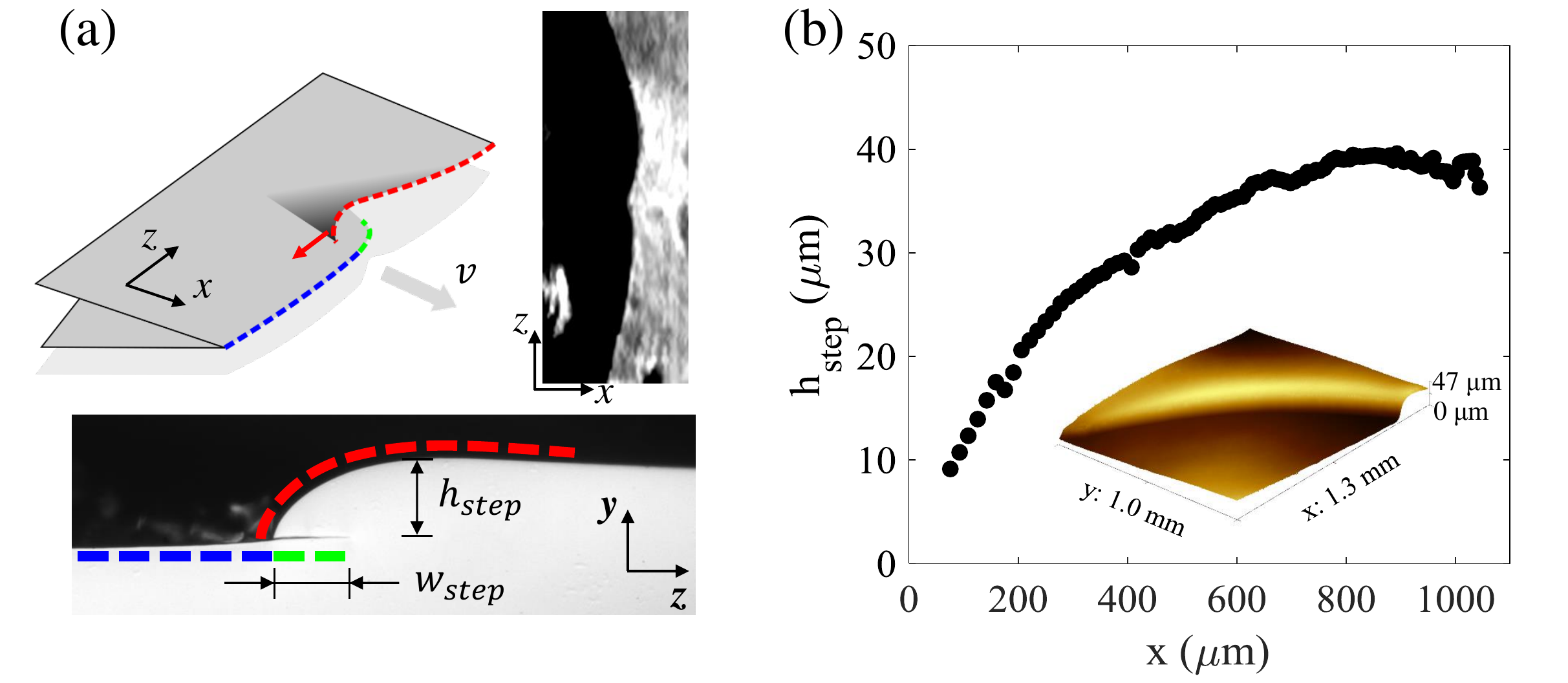}
\caption{\textbf{Topography of steps.} 
(a), Formation of steps. Top: Steps are formed by a discontinuous, disconnected crack front that contains an upper branch (red line) that, after some overlap (green line),  re-connects to the planar branch below (blue line) by sharply curving towards it. The step topologically retains its stability \citep{kolvin2018topological} as the planar branch always precedes the curved one.  (right panel) As noted in \citet{kolvin2018topological}, this configuration forms a cusp-like front, when projected on the $xz$ plane. A photograph of a typical cusp (by means of the configuration described in Fig.~\ref{fig1.fig}b) is presented. (bottom panel) A photograph in the $yz$ plane of a typical step formed within a fracture surface. The `hidden' overlapping section (green line), which lies beneath the curved branch (red line), is an extension of the planar branch (blue). Facets are formed as such steps progress along the $z$ direction in the direction normal to the curved section. (b), The measured step height, $h_{step}$, as a function of its front propagation distance along $x$ for a typical step. Once reflected, steps grow while propagating along the crack front. $h_{step}$  stabilizes at a height of about 40 $\mu$m \citep{kolvin2017nonlinear}. Inset: A 3D profilometer scan of a typical step.}
\label{fig6.fig}
\end{figure*}

Analysis of faceted cracks containing a single step was achieved using the experimental system described in Fig.~\ref{fig1.fig}b that enabled simultaneous measurement of the dynamics and structure of in-plane crack fronts. These were coupled with topographic measurements of the fracture surface formed by their propagation. Fig.~\ref{fig6.fig}a illustrates the topology of a step \citep{kolvin2018topological}. The crack front is composed of two disconnected and overlapping segments, a curved segment that partially overlaps a flat planar one. Both segments propagate simultaneously, while the flat segment is always slightly ahead of the curved one. Beyond the overlapping sections, the curved branch connects to the flat one and terminates. The overlapped section is hidden from view when viewing the fracture surface, but can be measured after sectioning the fracture surface along planes of constant $x$. A photograph of a $yz$ section of a typical step is presented in the lower panel of Fig.~\ref{fig6.fig}a, which clearly illustrates the different planes that comprise a step's structure. \citet{kolvin2018topological} demonstrated that each step generates a local increase in the total crack front extension of $\sim 1.4h_{step}$ that is formed by both the increase of the instantaneous height, $h_{step}$, on the fracture surface of the curved branch and the overlap width $w_{step}$, as shown in Fig.~\ref{fig6.fig}a. All of these contributions lead to increased local energy dissipation that, consequently, gives rise to in-plane deformation of the crack front profile; a local cusp-like shape  \citep{kolvin2018topological} as presented in Fig.~\ref{fig6.fig}a for a typical step, as it appears in our experimental system. 

\begin{figure*}[]
\centering
\includegraphics[width=1\linewidth]{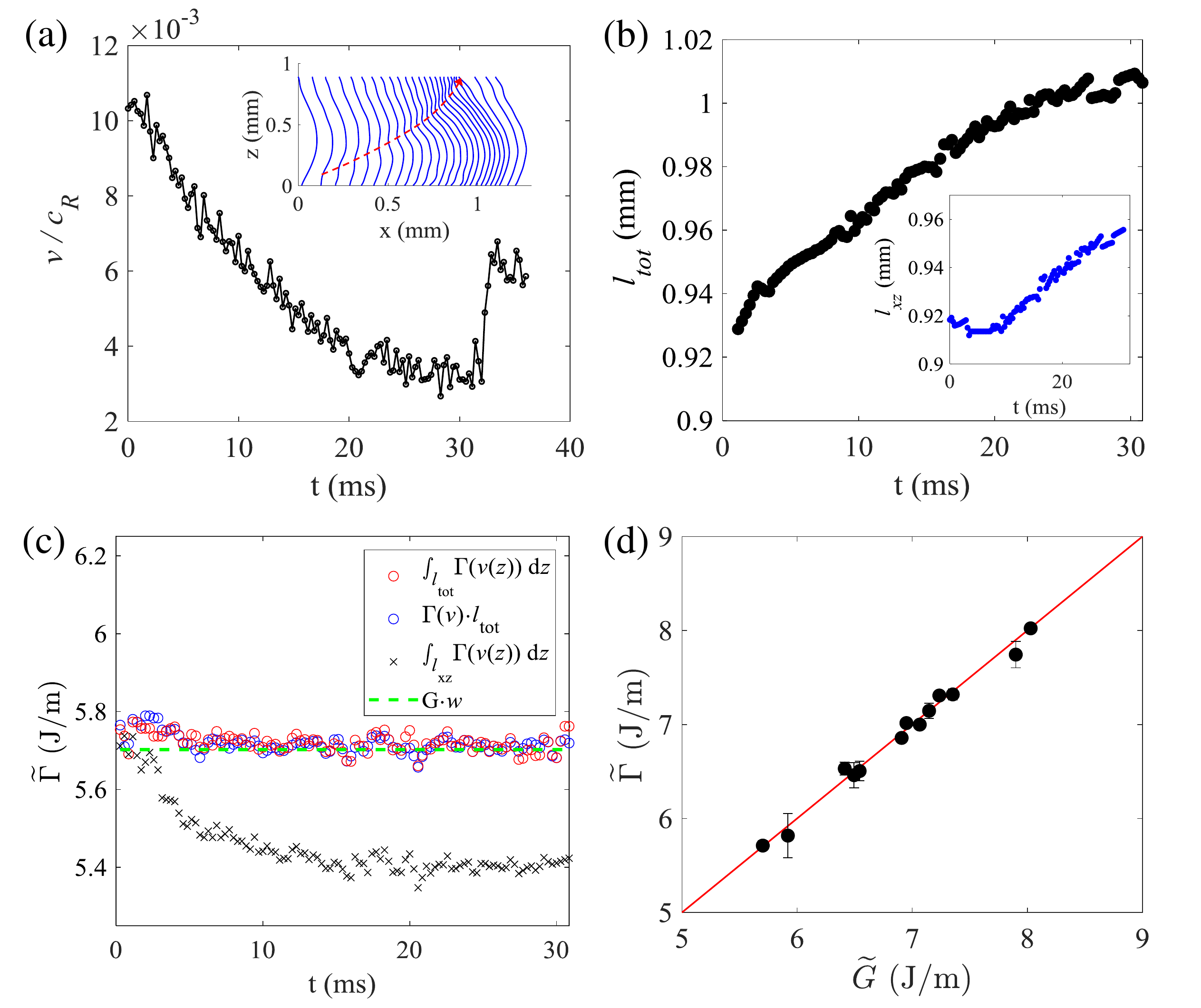}
\caption{\textbf{Energy balance of a faceted crack.} 
(a), The total crack front speed decreases with the growth of a step. The dynamics of the step presented in Fig.~\ref{fig6.fig}b is analysed. Inset: The corresponding sequence of instantaneous crack front shapes (displayed at 1.43 ms intervals) that are formed by the growing step ($0 < t < 36$ ms). (b), The total crack front length $l_{tot}\equiv l_{xz}$ + 1.4$h_{step}$, where $l_{xz}$ is the apparent front length, as projected onto the $xz$ plane.  $l_{tot}$ continuously grows as steps increase their height. Inset: The corresponding variation of the in-plane crack front length, $l_{xz}$. (c), The total fracture energy, $\widetilde{\Gamma}$, determined by $\int_{l_{tot}} \Gamma (v(z)) dz$ (red symbols), when taking into account the dynamics and all of the geometric variations of the crack front. Note that $\Gamma(v(z))$ is the fracture energy of {\it simple} cracks as presented in Fig.~\ref{fig3.fig}b, where we are using {\it local} velocities $v(z)$. $\widetilde{\Gamma}$ is invariant, and equal to the (constant) 3D energy flux $\widetilde{G}$ (green line) given by $G\cdot w$. The blue symbols correspond to the value of $\widetilde{\Gamma}$ calculated by the {\it mean} crack front speed $v$.
The black symbols represent the {\it apparent} fracture energy that would be obtained, were we to consider only the in-plane crack front length $l_{xz}$. (d), Energy balance between $\widetilde{\Gamma}$ and $\widetilde{G}$ for numerous different propagating cracks, each incorporating a single step. The red line represents $\widetilde{\Gamma}$ = $\widetilde{G}$. $G$ was calculated by means of Eq.~(\ref{eq-J.eq}), which remains valid for contours sufficiently far from the crack front.} 
\label{fig7.fig}
\end{figure*}

We measured the $h_{step}$ by means of profilometer scans of the fracture surface. The variation of $h_{step}$ along the fracture surface is shown in Fig.~\ref{fig6.fig}b. Upon reflection from one of the free faces of the sample (e.g. $z$ = 0), the step initiates from a height of about a few micrometers, grows as an approximate power law \citep{kolvin2017nonlinear}, and stabilizes at $h_{step} \sim$ 40 $\mu$m. The process generally repeats itself when the step encounters the other free surface (e.g. $z$ = $w$). When the sample surface is bounded by air, step reflection is commonly observed (see Fig.\ref{fig2.fig}b). Upon reflection, steps are inverted in orientation, lose height, again starting from a $h_{step}$ of a few microns, and propagate in the opposite direction. When the free surface is, however, bounded by water, the step is often not reflected, and disappears upon arrival at the free surface. We believe that this step `death' is the result of near-perfect acoustic transmission at the boundary between the gel and water.

Fig.~\ref{fig7.fig}a presents the detailed dynamics of a step-forming crack front as the step progresses from $z = 0$ to $z = w$. The inset of Fig.~\ref{fig7.fig}a shows the sequence of instantaneous in-plane ($xz$ plane) crack front shapes separated by a time interval of 1.43 ms. The step forms a locally concave front shape and travels along the crack front. Its location is highlighted by the red dashed arrow. The mean crack front speed $v$, determined by the average of the local crack speeds in $z$, is found to continually decrease with the step's growth. In this example, the free surface is bounded by a water bath, and the step `dies' when it impinges on the free surface. Upon the death of the step, the crack became a simple crack, with a typical curved front (see Fig.~\ref{fig4.fig}a) that propagates with a nearly constant speed. The decrease in the crack front speed is not the result of averaging in $z$; the step caused the entire crack front to decelerate.  This is revealed by the progressively decreased spacing between adjacent front positions presented in inset of Fig.~\ref{fig7.fig}a. Moreover, during the motion of the step, the whole curvature of the in-plane front shape continuously changed. Consequently, the length of the in-plane crack front, $l_{xz}$, continuously increased as $h_{step}$ grew (see inset of Fig.~\ref{fig7.fig}b). Using the measurement of $h_{step}$ (see Fig.~\ref{fig6.fig}b), the variation of the total crack front length $l_{tot}$, given by $l_{xz}$ + 1.4$h_{step}$, can be obtained, as shown in Fig.~\ref{fig7.fig}b.

With $l_{tot}$ in hand, we can now define the total energy dissipation of the entire crack front,  $\widetilde{\Gamma}$. This quantity, having physical dimensions $J\cdot m^{-1}$, is calculated by summing the contribution of local energy dissipation $\Gamma (v(z))$ for each point along the total extent, $l_{tot}$, of the crack front. This reads:
\begin{equation}
\widetilde{\Gamma} = \int_{l_{tot}} \Gamma (v(z)) dz\;,
\label{eq-Gamma3D.eq}
\end{equation}
where $\Gamma (v(z))$  is the fracture energy of a {\it simple} crack. $\Gamma (v(z))$ was determined using the local crack speed and the {\it independent} measurement of the fracture toughness provided in Fig.~\ref{fig4.fig}b. Since the overlapping structure of the step is coupled with the in-plane cusp of the crack front and moves at speed $v_{cusp}$, $\widetilde{\Gamma}$ in Eq.~(\ref{eq-Gamma3D.eq}) is given by $\int_{l_{xz}} \Gamma (v(z)) dz + 1.4h_{step} \cdot \Gamma (v_{cusp})$. The result of $\widetilde{\Gamma}$, measured at each instant shown in Fig.~\ref{fig7.fig}a, is presented in Fig.~\ref{fig7.fig}c (red symbols). The figure shows that the total energy dissipated by the whole crack front is invariant during the growth and motion of the step. Since the local normal speed $v(z)$ of the step-forming crack front weakly fluctuates around its mean value, $v$, we can also approximate $\widetilde{\Gamma}$ using the mean crack front speed: $\widetilde{\Gamma}\approx\Gamma (v)  l_{tot}$. As shown in Fig.~\ref{fig7.fig}c, the value of $\widetilde{\Gamma}$ calculated in this way is nearly indistinguishable from that calculated using the local crack front speed $v(z)$. The invariance of $\widetilde{\Gamma}$ is not trivial; it takes place despite the continuous development of the step, which significantly alters both its shape and crack dynamics. Furthermore, since $l_{tot}$ is correctly taken into account, the energy balance of the 3D crack front is retained; the value of $\widetilde{\Gamma}$ is precisely equal to the total energy flux into the crack front $\widetilde{G} = G \cdot w$, where $G$ represents the energy flux far away from the crack, determined by the speed of the steady-state propagation of the crack using Fig.~\ref{fig3.fig}b.

For comparison, we also present the energy dissipation that is obtained were we not to consider the increased crack lengths induced by the step in Fig.~\ref{fig7.fig}c. The discrepancy with $\widetilde{\Gamma}$ underscores the fact that all of the variations in crack dynamics and crack geometries must be taken into account. Without accounting for all of the crack front variations imposed by a step, the constancy of $\widetilde{\Gamma}$ for constant $G$ would not be apparent,  for the typical example presented in Fig.~\ref{fig7.fig}c. More generally, the equality of $\widetilde{\Gamma}$ and $\widetilde{G}$ is shown to be generally valid in Fig.~\ref{fig7.fig}d, where $\widetilde{\Gamma}$ is compared with $\widetilde{G}$ for numerous different step-forming cracks.

\section{Discussion}

Special attention must be drawn when any dissipative mechanism breaks the invariance in $z$ that is necessary for effectively 2D behavior. Upon any change of the local dissipation, crack fronts may undergo significant changes in dynamics, geometry, and even topology, while still retaining global energy balance. A particular example is the presence of steps; conserving energy balance while not being described by 2D LEFM. Examples of this can be seen in the sharp jumps of the local velocity despite a constant $G$, both upon step reflection from free boundaries (Fig. \ref{fig2.fig}) and the transition from stepped cracks front to simple cracks (e.g. Fig. \ref{fig7.fig}a).  Below we discuss a number of phenomena that have been clarified by this study.

\subsection*{Incorporating crack front structure for slow cracks:}
Once the `bare' fracture energy $\Gamma(v)$ is known (e.g. Fig. \ref{fig3.fig}b), our results suggest, at least for slow fracture in 3D isotropic materials, that geometric considerations of crack fronts are all that is required for calculating the fracture energy. The opposite is also true; if one performs such a comparison, and empirically finds that energy balance is not conserved, then there is a strong likelihood that some aspect of the geometrical crack front structure has been missed.  \citet{tanaka2000fracture} understood this point and incorporated crack surface roughness of faceted cracks in their estimates for  $\Gamma(v)$. As Fig. \ref{fig7.fig} however shows, simply incorporating the surface roughness is insufficient. The true determination of $\widetilde{\Gamma}$ depends critically on the details of the crack front state; both the crack's out-of-plane structure as well as its curvature and length must be taken into account. Without doing this, any apparent `characteristic' dissipation that would result from naive 2D energy balance would lead to significant errors. Such misinterpretation could lead to discrepancies or effective `state dependence' in perceived values of $\Gamma(v)$ as well as `effective' dissipation of the 2D problem that are inconsistent with crack dynamics (as could be interpreted from the in-plane calculation in Fig.~\ref{fig7.fig}c). 

\subsection*{ Incorporating crack front structure for dynamic cracks:} Once a crack becomes strongly dynamic (e.g. $v>0.5C_R$) then inertial effects become important and purely geometrical contributions to $\widetilde{\Gamma}$ must be supplemented by dynamic ones. In this vane, \citet{sharon1999confirming}  showed that, for rapid dynamic cracks undergoing micro-branching (and consequent crack speed fluctuations), the application of energy balance to predict the mean crack speed (even while incorporating all of the additional surface created by the extensive micro-branches in $\Gamma(v)$) is not sufficient to determine the equation of motion for the resultant mean crack velocity. This same study showed that only the crack speed of instantaneously {\it simple} crack states (for which the inertial contributions are correctly incorporated by LEFM) could correctly evaluate $\Gamma$($v$) of the material.

\subsection*{Local Energy Balance:} In this work, we have measured the local energy dissipation $\Gamma(v(z))$ along the crack front and the  total energy flux $G$ into the full crack front to probe the global energy balance $\widetilde{\Gamma}$ = $\widetilde{G}$. While we have not directly  demonstrated {\it local} energy balance, we believe that this property can  be inferred from our measurements.  Our experiments have demonstrated that energy balance is maintained in {\it each} moment in time, regardless of the instantaneous crack front length or step position and amplitude. The result that global energy balance holds for the numerous arbitrary and continuously varying crack front shapes and dynamics that were sampled, therefore, constitutes a proof that {\it local} energy balance $\Gamma (v(z)) = G (v, z)$ at each point along the crack front also takes place. This result has been implicitly assumed in many previous studies, when non-planar fronts have been considered, but this assumption had never been explicitly verified. Local energy balance was suggested by the work of \citet{chopin2011crack}, who used Gao and Rice's first-order perturbation analysis of the crack front \citep{Gao1989FirstOrder} to demonstrate that local energy balance provides a good explanation of the crack front profile when a crack  moves along the boundary formed by a manufactured jump in fracture energy. \citet{kolvin2018topological} later used this analysis to quantitatively describe the shape of a crack front resulting from  a propagating step. 

\subsection*{Generality of the results:} 
Here, we have focused on the overall structure of stepped crack fronts in polyacrylamide gels. The structure and influence of these steps may be similar to the mechanisms that give rise to observed faceted fracture surfaces in other brittle amorphous materials.  Well known phenomena include `lance-like' or `twist-hackle' structures in glass \citep{sommer1969formation, hull1999fractography}, and faceted fracture surfaces in gelatin \citep{baumberger2008magic,pham2016growth} and Homalite H-100 \citep{pham2014further, pham2016growth}. Facets in such amorphous materials should be qualitatively different than the faceted fracture surface in brittle crystalline materials, as the latter is formed by deflected crack fronts propagating along multiple crystal planes \citep{kermode2008low}. Despite any differences in the physical natures of facets, the necessity of accounting for crack front structure is generally true in any analysis of energy balance. For example, we have seen in Fig. \ref{fig4.fig} that even for the `trivial' case of simple cracks, the crack front structure (crack front curvature) should be taken into account to provide precise values for $\Gamma$. 
\\

While this work has clarified much of the influence of secondary structure on crack front dynamics, many questions still remain. Below, we note a number of important unresolved issues.

\subsection*{Stability and Bistability:} We have found that the formation of simple cracks at low speeds critically depends on the imposed initial and loading conditions. Once formed, simple cracks remain surprisingly stable. On the other hand,  at `zero' velocity, very slight anti-planar perturbations of the initial crack or slight mode mixity in the loading will cause simple cracks to lose stability and  excite stepped cracks. Once excited, the stability of steps in gels is maintained, as \citet{kolvin2018topological} have shown, because of the step topology. At intermediate velocities (e.g. $v\sim 0.1c_R$) simple cracks become, apparently, much more stable than faceted cracks; empirically, stepped fronts generally transition to simple cracks for sufficiently large $v$. Thus, bistability is lost. The mechanism driving this transition is still unknown.

\subsection*{Anisotropic Materials:} Determination of the fracture energy using solely geometric considerations may not be true for ductile materials like anisotropic metallic alloys \citep{pineau2016failure}. The crack fronts in these materials are usually complex and even subtle details of the mode of fracture (e.g. plane strain vs. plane stress) may well give rise to different values of the fracture energy (via e.g. the local selection of different local fracture planes). These materials may therefore undergo complex and stress-direction related dissipative mechanisms \citep{garrison1987ductile} that are more complex than the simple geometrical considerations that hold for isotropic materials, such as gels. Such complexity could  also invalidate the measure of $G$ by the use of $J$-integral measurements performed on a single material plane or free surface, such as those utilized in our experiments.

\subsection*{Validity of the 2D J-integral to 3D systems:} Even in effectively 2D systems, the accuracy of the $J$-integral is based on the fact that all of the dissipative mechanisms are confined to the near-tip region. This, for example, may not be the case when mechanisms such as poro-viscosity are significant \citep{baumberger2006solvent, noselli2016poroelastic}. In such cases, where significant bulk dissipation occurs \citep{bouklas2015effect, yu2018steady}, path {\it dependence} will yield non-trivial information about material properties. In our case, the path independence of our measurements (see Fig.~\ref{fig3.fig}a) implies, for example, that (for the composition of polyacrylamide gels that we used) poro-viscous contributions can indeed be neglected.

For non-simple crack fronts, if a contour approaches the crack front too closely, the 2D $J$-integral will no longer be an accurate measure of the strains surrounding the crack front as translational invariance along $z$ is lost. Care should then be taken to ensure that interpretations obtained by such 2D calculations are valid; calculations must only be performed at distances such that the crack front structure does not significantly affect the $z$ variation of the fields far from the crack front.
\\

\subsection*{Properties of Steps:} Many properties of crack front steps remain unresolved. These include how their steady-state height is determined, which for the gels used in these experiments is $h_{step}\sim$ 40 $\mu$m. In these polyacrylamide gels, steps always grow upon nucleation or decay upon merging until stabilizing at $h_{step} \sim$ 40 $\mu$m. What is the significance of this 40 $\mu$m scale? \citet{kolvin2018topological} conjectured that once a step emerges, the curved section, by generating a small anti-planar perturbation, produces a significant repulsion from the straight branch \citep{melin1983cracks, schwaab2018interacting}. As branch separation should lead to a decrease in repulsion strength, it was conjectured that the separation distance of the two branches forming a step should stabilize at the distance corresponding to when the repulsion balances the attraction. As such, the stable height of the step may be an intrinsic property, independent of the material properties, but dependent on a scale such as the size of the overlapping section between the two fracture planes that form a step. This, however, has yet to be demonstrated and, of course, is related to the complex 3D spatial structure of crack front steps, which itself remains a theoretical/numerical challenge.

\subsection*{What determines the dynamic behavior of $\Gamma$?}  Our measurements of $\Gamma$ for {\it simple} cracks even in {\it `simple'} isotropic materials such as the elastomers used here, raise additional questions about the role of the internal structure of polymers during the fracture process.  For example, why do elastomers have such a strong and monotonically increasing fracture energy dependence for low values of $v$? One might expect that the opposite would take place; prior to any crack extension due to fracture of the material, the tangled polymer chains that make up a gel, if given sufficient time (small $v$) should undergo large elongation and alignment, as well as internal friction of the polymer strands \citep{yang2019polyacrylamide, baumberger2020environmental}. In this picture, many of these dissipative processes would, conceivably, not have time to develop at high values of $v$, so that naively one might expect the fracture energy to {\it decrease} with $v$. As Figs. \ref{fig3.fig} and \ref{fig4.fig} demonstrate, this is obviously a wrong (or, at least, incomplete) picture. A quantitatively accurate description of how this class of materials does break, poses (in our view) an interesting challenge. If the large growth of $\Gamma$ is due, in some way to the internal structure of the elastomers used here, one might expect to obtain the same behavior for other polymers as a function of $v$. While this is an important question, there is, currently, insufficient information to perform such a comparison. 
\\

In conclusion, we have shown here that energy balance is indeed valid but {\it only} when all of the geometric and dynamic variations of the 3D crack front are quantitatively accounted for. Thus, hidden structure can, indeed, trigger unexpected consequences and, often counter-intuitive, dynamic behavior. The results of this study may have numerous implications to both our fundamental understanding of fracture and resulting material properties such as fracture toughness. We have shown that even for the very `simple' case of the fracture of a brittle material at quasi-static speeds (where inertial effects are negligible), internal structure of cracks (or their internal `state') will play a crucial role in determining both fracture dynamics and `effective' fracture toughness. Realizing this is critically important to our interpretation of observations in seemingly simple physical situations.

\section{Acknowledgement}
M.W. and J.F. gratefully acknowledge the support of the Israel Science Foundation, grant $\#$840/19. This work was supported by the International Research Project ‘Non-Equilibrium Physics of Complex Systems’ (IRP-PhyComSys, France-Israel). M.A.-B. and M.W. acknowledge the support of the Lady Davis Fellowship Trust. J.M.K acknowledges SNSF grant no. 200021$\_$197162.

\bibliographystyle{elsarticle-harv} 
\bibliography{cas-refs}





\end{document}